\documentclass[preprint,3p,times,twocolumn,authoryear]{elsarticle}
\usepackage[colorlinks=true, allcolors=blue]{hyperref}
\usepackage[utf8x]{inputenc}
\usepackage{graphicx}
\usepackage{framed} % Framing content 

\usepackage{amssymb}
\usepackage{amsthm}

\usepackage[usenames]{color} %used for font color
\usepackage{amsmath} %maths

\usepackage{color,soul}
\usepackage{longtable}
\journal{ArXiv}

\begin{document}
	\begin{frontmatter}
		
		\title{A Bayesian approach to recover the theoretical temperature-dependent hatch date distribution from biased samples: the case of the common dolphinfish (\textit{Coryphaena hippurus})\tnoteref{CC_license}}
		
		\tnotetext[CC_license]{© 2020. This manuscript version is made available under the CC-BY-NC-ND 4.0 license \href{http://creativecommons.org/licenses/by-nc-nd/4.0/}{http://creativecommons.org/licenses/by-nc-nd/4.0/}}
		
		%% Include affiliations in footnotes:
		\author[IMEDEA]{Vicenç Moltó}
		\author[IMEDEA]{Andres Ospina-Alvarez \corref{mycorrespondingauthor}}
		\cortext[mycorrespondingauthor]{Corresponding author}
		\ead{aospina.co@me.com}
		% \ead[url]{www.andresospina.eu5.net}
		\author[MALTA]{Mark Gatt}
		\author[IMEDEA]{Miquel Palmer}
		\author[IMEDEA]{Ignacio A. Catalán}
		
		\address[IMEDEA]{Mediterranean Institute for Advanced Studies IMEDEA (UIB-CSIC), C/ Miquel Marques 21, CP 07190 Esporles, Balearic Islands, Spain.}
		\address[MALTA]{Department of Fisheries and Aquaculture, Fort San Lucjan, Triq il-Qajjenza, Malta.}
		
		\begin{abstract}
			
	Reproductive phenology, growth and mortality rates are key ecological parameters that determine population dynamics and are therefore of vital importance to stock assessment models for fisheries management. In many fish species, the spawning phenology is sensitive to environmental factors that modulate or trigger the spawning event, which differ between regions and seasons. In addition, climate change may also alter patterns of reproductive phenology at the community level. Usually, hatch-date distributions are determined back-calculating the age estimated on calcified structures from the capture date. However, these estimated distributions could be biased due to mortality processes or time spaced samplings derived from fishery. Here, we present a Bayesian approach that functions as a predictive model for the hatching date of individuals from a fishery-dependent sampling with temporal biases. We show that the shape and shift of the observed distribution is corrected. This model can be applied in fisheries with multiple cohorts, for species with a wide geographical distribution and living under contrasting environmental regimes and individuals with different life histories such as thermo-dependent growth, length-dependent mortality rates, etc. 

		\end{abstract}
		
	\end{frontmatter}
	
	\section{Introduction}
	
	Reproductive phenology, growth and mortality rates are key ecological parameters that determine population dynamics and are therefore of vital importance to stock assessment models for fisheries management \citep{Lee:2011bl, Methot:2013dm,Stawitz:2019kj, Walker:2019ce}. These parameters are difficult to estimate and are not known for all fisheries, or even applied in the stock assessment models, especially in data-poor fisheries. This fact could lead to considerable sources of bias for the stock assessment models.
	
	The  spawning phenology  is sensitive to environmental factors that modulate or trigger the spawning event, which differ among regions and seasons \citep{Portner:2010ega, Poloczanska:2016kka}, climate change can also alter the patterns of reproductive phenology at the community level \citep{Pankhurst:2011hga, Anderson:2013ho}. If optimal condition (food availability, lower depredation rates) are observed at the beginning of the spawning season, growth and condition is early promoted and survival to juvenile stages is increased \citep{Lapolla:2005bl, Islam:2015ej}. On the contrary, when optimal environmental factors occur later in the spawning season, mortality rates are considerable (e.g., higher accumulated depredation pressure, stronger effects of advective or dispersive processes). In this situation, being born later in the spawning season could be advantageous \citep{Yoklavich:1990ve, McGovern:2011bl, Xie:2005ih, Folkvord:2015gn}. One way to determine the hatching date distribution is back-calculating the age from hard structures of fish (e.g., otoliths captured in a fishery or scientific survey). However, the inferred distribution of observed birth-dates distribution could be incorrect  due to sampling time bias. The earlier a fish is born, the more likelihood there is that it will die before the sampling date \citep{Campana:1992wb}. Fishery-dependent sampling typically suffers from such a temporal bias in the observed hatch-day distribution because fish are vulnerable only after reaching a certain size/age threshold or simply due to the existence of a mandatory beginning of the fishing season. Accordingly, the spawning probability can only be properly estimated from the observed birthday distribution after incorporating the effect of mortality rates. 
	
	We present here a Bayesian approach that functions as a predictive model for the hatching date of a population that is sampled with a temporal bias. Our model for predicting the hatching date can be applied in fisheries with multiple cohorts, for species with a wide geographical distribution and living under contrasting environmental regimes and individuals with different life histories (e.g. thermodependent growth, length-dependent mortality rates, etc.). Therefore, the work presented here is relevant to fisheries management, ecology and biology. The model can be incorporated into habitat prediction models, individual-based growth models, bio-energetic models, models projecting the impact of climate on fish growth, etc. In this example, we apply the model to the dolphinfish (\textit{Coryphaena hippurus}) fishery in the NW Mediterranean \citep{CopeMedII:2016vc}. The fishery is based on age-0 individuals that are spawned in summer in the area and are fished between August and December \citep{CopeMedII:2016vc} after an extremely fast growth \citep{Massuti:1999jd}.
	
	\section{Methods}
	Three main data sources were required to assess the seasonal pattern of spawning probability of dolphinfish: DB1. A database of juveniles’ daily otolith readings; DB2. A database of gonadosomatic index (GSI) estimations  (indicative of average population spawning state) around the world; and DB3. Sea Surface Temperature (SST) estimations extracted from satellite imagery for each location and date sorted in DB2.
	
	The rationale behind coupling these three databases respond to: 1) The hatch date distribution of the sampled fish should be related with the hatch date distribution of the cohort \citep{Methot:1983eb}; 2) The seasonal pattern of GSI should be a proxy of the seasonal spawning pattern \citep{McQuinn:1989ja}; 3) Fish maturity and GSI are strongly related with the thermal characteristics of their spawning habitat \citep{Dobson:1977gu,Neuheimer:2014ja, Pankhurst:2011hga}.
	
	Therefore, two sub-models can be built: The first one predicts the probability $\operatorname{\emph{Prob. birth}}_{i, j}$ that an individual $i$ belonging to the cohort $j$ was be born on the date $\operatorname{\emph{hatch}}_{i, j}$; and the second one to predict the fraction of the population being at the spawning state $\operatorname{\emph{Prob. GSI}}_{t,j}$ at the month $t$. Both sub-models are related to the temporal location of the spawning peak ($\mu_{j}$) and the spread of the distribution of the actual dates of birth ($\sigma_{j}$).
	
	\begin{framed}
	\textbf{Box 1. Databases} \\
	\\
	\underline{DB1. Juveniles’ otolith readings database} \\
	1074 individuals, with recorded capture date,  corresponding to 13 Mediterranean cohorts (3 years from Balearic Islands and 8 years from Malta; Table \ref{tab:table-1}).\\ 
	\\
	\underline{DB2. Monthly population-averaged GSI database} \\
	\ Population-averaged GSI and date for 7 populations  along a wide latitudinal gradient (Taiwan, Malta , Balearic Islands, North Carolina, South West India,  
	Japan and California; Table \ref{tab:table-2})\\ 
	\\
	\underline{DB3. Monthly averaged SST} \\
	SST from satellite imagery for the locations (averages of 100x100 km around the areas of capture) and dates  from DB2 (Not shown here).\\
	\end{framed}
	
	\subsection{Prob. birth$_{i, j}$ sub-model}
	The hatch-date (birth) distribution sub-model is derived with the assumption of the hatch dates of a given cohort are normally distributed in time around a cohort-specific ($j$) peak date ($\mu j$; Julian days from January 1st), and a cohort-specific spread ($\sigma j$; days). The fish $i$ has been actually captured at a given date and has a known age and the probability can be approximated by:
	
	\begin{multline}\label{Eq.1}
	\operatorname{Prob. birth}_{i, j}=\\
	\frac{\operatorname{dnorm}\left(\operatorname{hatch}_{i, j},  \operatorname{mean}=\hat{\mu}_{j}, \operatorname{sd}=\sigma_{j}\right)}{ \operatorname{pnorm (capture}\left._{i, j}-\operatorname{age-min} _{i, j},  \operatorname{mean}=\hat{\mu}_{j},  \operatorname{sd}=\sigma_{j}\right)}\
	\end{multline}
	
	where $dnorm$ denotes the probability density function and $pnorm$ denotes the cumulative probability function of the normal distribution; $hatch_{ij}$ is the observed (inferred from the otolith) hatch date of the $i$ fish from the $j$ cohort; $capture_{ij}$ is the capture date (Julian days); $\operatorname{\emph{age-min}}_{ij}$ is the age from which the fish $i$ reaches a threshold size after which it is considered vulnerable to fishing (20 cm; which is the approximate average length of the minimum size at catch in the available fishery-dependent data series (DB1)); and therefore, when the fishing mortality acts. This age is estimated using a growth model developed for this species (Moltó et al., submitted). Finally, $\hat{\mu}_{j}$ represents the apparent spawning peak after being distorted by natural and fishing mortality, which is given by:
	
	\begin{equation}\label{Eq.2}
	\hat{\mu}_{j}=\mu_{j}+m \sigma_{j}^{2}
	\end{equation}
	
	where $\mu_{j}$ is the true peak location (Julian days) of the cohort $j$ and $m$ is the instantaneous natural mortality rate ($1.703 \operatorname{day}^{-1}$), estimated from the HoenigNLS empirical prediction model using the maximum known life span (Tmax) for a given species  as it was the best empirical estimator of natural mortality, according to a study of more than 200 species \citep{Then:2014bq}. Here, Tmax was set as 3 for dolphinfish \citep{Massuti:1999jd}.
	
	At the between-cohort level, the spawning peak location ($\mu_{j}$) and the spread ($\sigma_{j}$) are assumed to be a linear function of the sinusoidal temperature profile of a given year and region:
	
%	\begin{multline}\label{Eq.3}
%	\mu_{j}=\beta_{o} + \beta_{1} \text {Temp. Mean}_{j} + \beta_{2} \text {Temp.  Amp}_{j} +\\ \beta_{3} \text {Temp. Phase}_{j} + \varepsilon_{j}
%	\end{multline}

	\begin{multline}\label{Eq.3}
	\mu_{j}=\beta_{o} + \beta_{1} \text {Temp. Phase}_{j} + \varepsilon_{j}
	\end{multline}
	
%	\begin{multline}\label{Eq.4}
%	\sigma_{j} = \gamma_{o} + \gamma_{1} \text {Temp. Mean}_{j} + \gamma_{2} \text {Temp. Amp}_{j} +\\ \gamma_{3} \text {Temp. Phase}_{j}
%	\end{multline}
	
	\begin{multline}\label{Eq.4}
	\sigma_{j} = \gamma_{o} + \gamma_{1} \text {Temp. Mean}_{j}
	\end{multline}
	
	where $\epsilon_{j}$ is either a normally distributed error with zero mean and standard deviations $\text{\emph{sd}}. \mu$ (Eq. 3) or $\sigma_{j}$ is assumed to be gamma distributed with a rate $\text{\emph{sd.rate}}$ (Eq. 4). $\text{\emph{Temp.Mean}}_{j}$, $\text{\emph{Temp.Amp}}_{j}$ and $\text{\emph{Temp.Phase}}_{j}$ are the mean, amplitude and phase of a sinusoidal function fitted to the temperature profile that has experienced the cohort $j$:
	
	\begin{multline} \label{Eq.5}
	\text {Temp}_{t, j} = \text {Temp. Mean}_{j}+\text {Temp. Amp}_{j}\\ \sin \left(2 \pi t / 365 + \text {Temp. Phase}_{j}\right) + \varepsilon_{j}
	\end{multline}

	\subsection{Prob. GSI$_{i, j}$ sub-model}
	The spawning state (GSI) distribution sub-model is derived from the gonadosomatic index data. The observed $\text{\emph{GSI}}_{t,j}$ (average GSI of a sample of fish from the population $j$ at the month $t$, in Julian days) is modelled by:
	
	\begin{equation}\label{Eq.6}
	\operatorname{\emph{GSI}}_{t, j} = \operatorname{\emph{GSI max}}_{j} \operatorname{\emph{Prob. GSI}}_{t, j} + \varepsilon_{j}
	\end{equation}
	
	where $\operatorname{\emph{GSI max}}_{j}$ is the maximum GSI value attainable at the population $j$ (scale factor); $\operatorname{\emph{Prob. GSI}}_{t,j}$ is the fraction of the population being at the spawning state at the month $t$; and $\epsilon_{j}$ is a normally distributed error with zero mean and standard deviations $\operatorname{\emph{sd. GSI}}$. In turn, $\operatorname{\emph{Prob. GSI}}_{t,j}$ is related with the spawning peak location ($\mu_{j}$) and the spread ($\sigma_{j}$) of the actual hatch dates distribution (common parameters shared with $\operatorname{\emph{Prob. birth}}_{i, j}$ sub-model) by a super Gaussian distribution model \cite{Decker:1994wo}, that only differs with the normal distribution by the exponent $\operatorname{\emph{POW}}$.  When $\operatorname{\emph{POW}}<1$, the resulting distribution has a pointed peak and long tails ("Christmas tree distribution").
	
	\begin{equation}\label{Eq.7}
	\operatorname{\emph{Prob. GSI}}_{t, j} = \frac{1}{\sigma_{j} \sqrt{2 \pi}} \mathrm{e}^{\left(-\frac{\left(t-\mu_{j}\right)^{2}}{2 \sigma_{j}^{2}}\right)^{\operatorname{\emph{POW}}}}
	\end{equation}
	
	Therefore, eq. 7 effectively links the (unbiased) probability of spawning with the observed temperature profile, and the relationship between GSI and temperature. The parameters of the two sub-models above have been estimated using a Bayesian approach implemented in a custom R script \citep{RCoreTeam:2019wf} (Appendix 1) that runs JAGS \citep{Plummer:2003wk} for moving the MCMC chains. We use the "ones trick" \citep{Kruschke:2015ti} for sampling from the non-standard distribution described in Eq. 1. Three independent chains were run. The convergence of the MCMC chains was assessed by visual inspection of the chains and was tested using the Gelman-Rubin statistic \citep{Plummer:2008ho}.
	
	A threshold value of 1.1 or less was assumed to suggest convergence \citep{Gelman:2013tc}. Posterior distribution was estimated by at least 3 000 valid iterations after appropriate burning and thinning (one out 10 iteration were kept). The first 10 000 iterations were not considered. Nearly uninformative priors have been assumed for all the parameters. Specifically, priors for  $\beta$ (Eq. 3) and $\gamma$ (Eq. 4) parameters are assumed to be normally distributed with zero mean and tolerance $= e^{-10}$, GSI.scale are assumed to be normally distributed with zero mean and tolerance $= e^{-10}$, but constrained to be positive, and all tolerances and rates (\emph{sd. rate} in Eq. 4) are assumed to be gamma distributed with parameters shape$=0.01$ and scale$=0.01$.
	
	Finally, to encourage the reproducible results, we make our code public on a GitHub repository
	\href{https://doi.org/10.5281/zenodo.3725530}{doi:10.5281/zenodo.3725530}
	
	\section{Results}
	
\begin{figure*}
		\centering
		\includegraphics[width=0.7\linewidth]{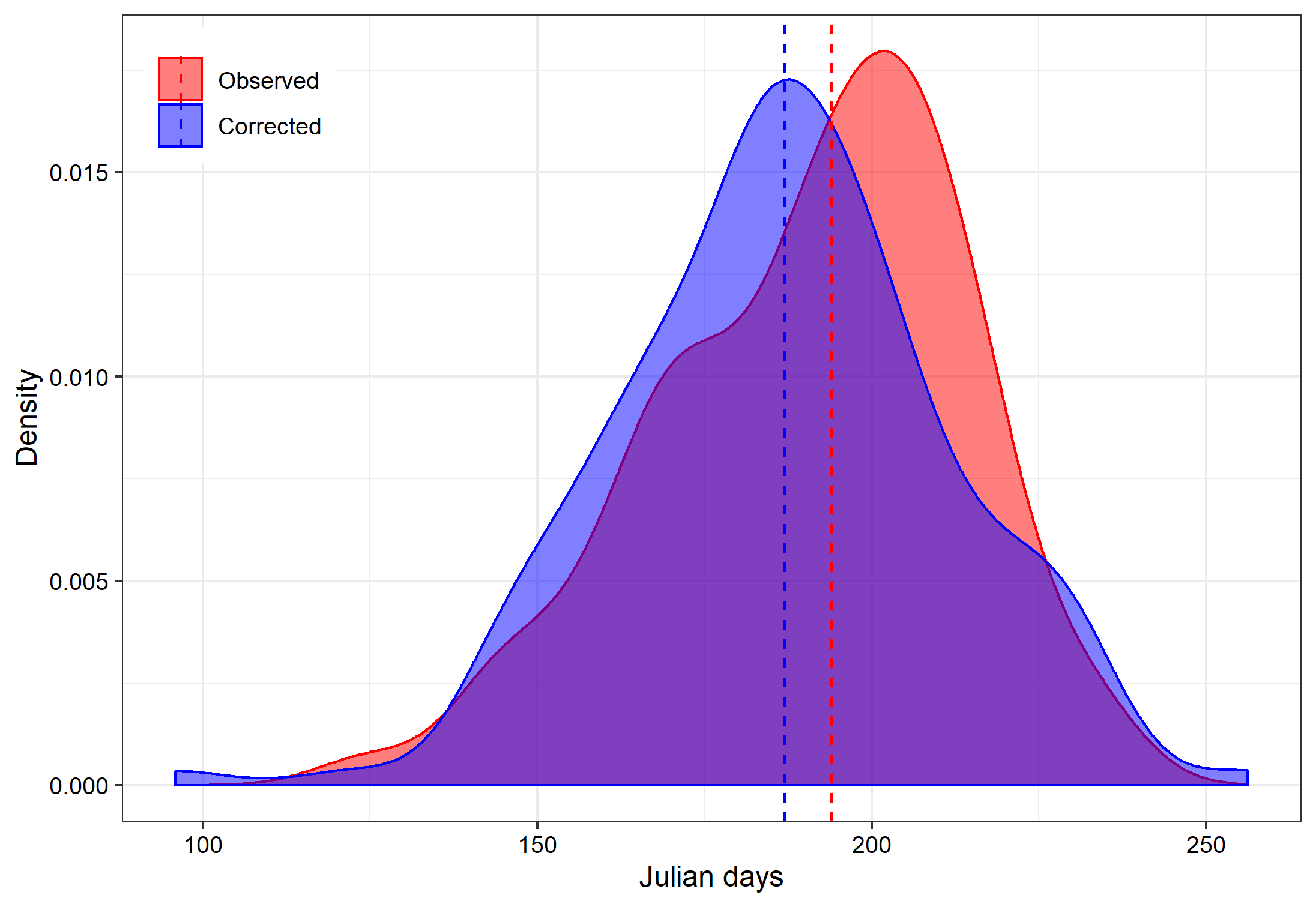}
		\caption{Shift in hatching date distribution for a dolphinfish cohort from Balearic Islands for the year 2004. The red shaded area corresponds to the distribution of observed hatching dates and the blue area represents the distribution of theoretical hatching dates estimated by the Bayesian method explained in this research. The vertical bars (red and blue) indicate the mean values of each distribution. 
		}
		\label{fig:figure-1}
	\end{figure*}

\begin{figure*}
	\centering
	\includegraphics[width=0.7\linewidth]{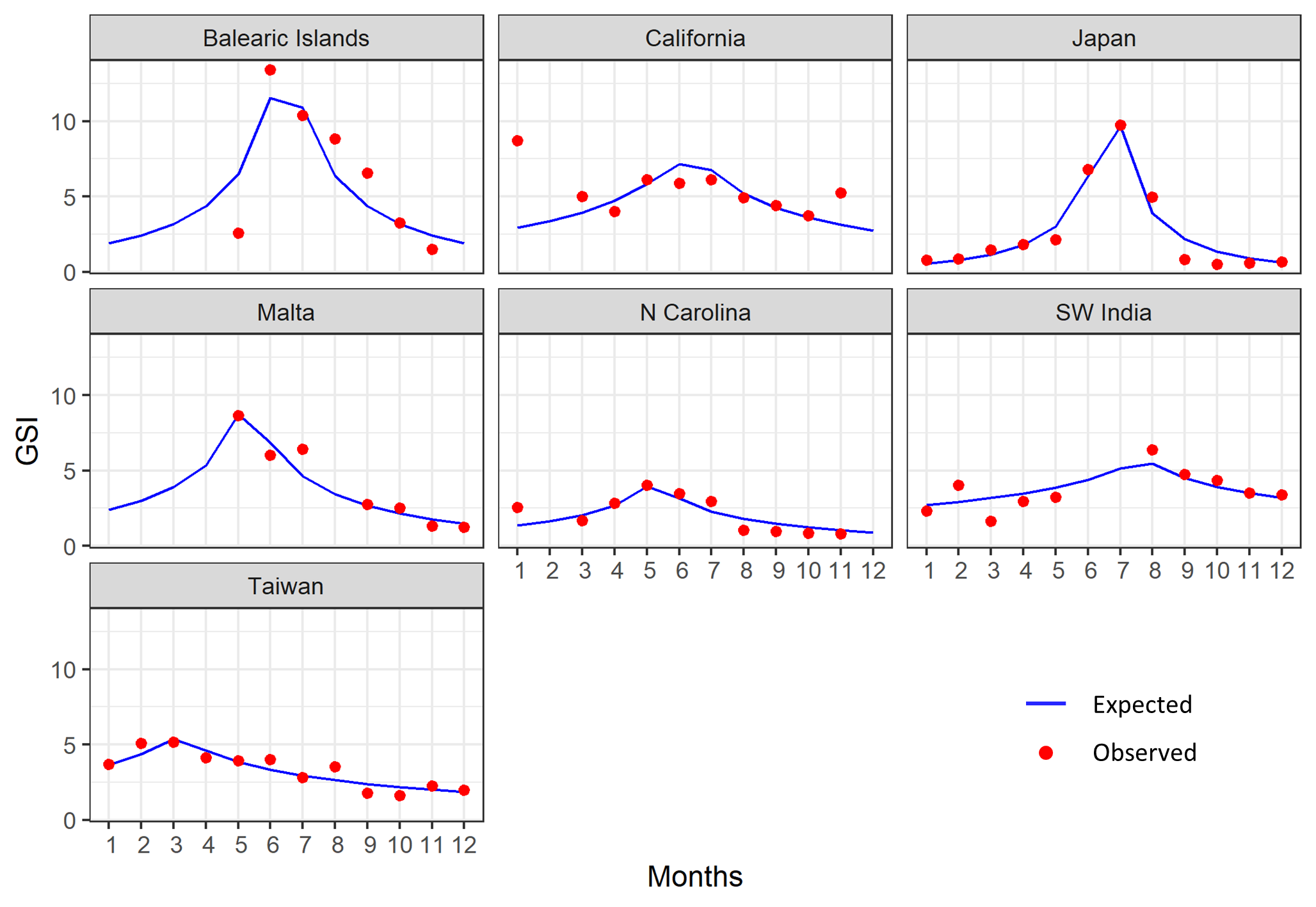}
	\caption{GSI patterns of the dolphinfish populations examined. 
	}
	\label{fig:figure-2}
\end{figure*}

The theoretical (corrected) hatch-date distribution has been estimated for the Mediterranean dolphinfish cohorts above mentioned (DB1). As an example, the corrected hatch-date distribution for the Balearic Islands 2004 cohort is shifted to the left (early in the spawning season) with respect to the observed hatch-date distribution, with a difference of 5 days in their median values (Fig. \ref{fig:figure-1}). Our model also shows an adequate reproducibility for the GSI patterns of the populations examined (DB2), which is shown in the Figure \ref{fig:figure-2}.

	\section{Discussion}
	
	Daily growth based on the analysis of the otolith microstructure, together with information from the gonads, has been widely used to obtain information on the reproductive biology of fish species.  From ichthyoplankton studies, or based on later-stage individuals, hatching date distributions, determined by backward calculations of age from the date of capture in otolith readings, is an extended methodology to determine spawning behaviour and, therefore, environmental factors related to it  \citep{Campana:1992wb}. However, the hatching date distribution obtained from these observations could be biased due to the lack of individuals at certain stages, mainly due to advective or dispersive processes or to mortality rates that occur during the time between the birth of the individuals and their capture \citep{Campana:1992wb}. The  majority of studies that have determined hatching date distributions using this methodology are based on larvae or early juvenile (young of the year) stages, where the correction factors applied are based on the absence or survival of individuals between successive cohorts or age classes (most of which are arbitrarily set within the same spawning season). For example, \cite{Methot:2013dm} corrected the hatch-date distributions by the inverse of survival rates between age classes, whereas others studies, focused on larval stages, corrected the hatch distribution through the instantaneous mortality rate determined for the life period studied \citep{Yoklavich:1990ve, Fortier:1998fm, Marteinsdottir:2000kq, QuinonezVelazquez:2000hy}. \cite{QuinonezVelazquez:2000hy} also applied a correction based on a mortality coefficient for juvenile pacific sardine. All these studies are conducted within the critical life period for fish species, when mortality rates are highly significant, with strong effects in the shape of the back-calculated distributions \citep{Campana:1992wb}. These authors affirm that after the critical period, once juvenile have attained a determinate age, in which mortality rates have decreased to minimum values, the shape of the observed and expected hatch-distributions will be similar. This is the case of our study, because although we are sampling on young of the year individuals, they are already survivors of the critical early life period. Thus, we applied a constant natural mortality coefficient acting throughout the entire life period. Moreover, once individuals attain a certain length (with is variable in age depending on the thermal history experienced according to the growth model used (Moltó et al. submitted), they will be available (vulnerable) to the fishery which is, in turn, legally established with a certain period in the Mediterranean Sea (Recommendation GFCM/30/2006/2 available online at: \href{http://www.fao.org/gfcm/data/reporting/dolphinfish/en/}{http://www.fao.org/gfcm/data/reporting/dolphinfish/en/}). Once the fish is vulnerable to the fishery, the fishing mortality could act in a similar way to the natural mortality during the critical stage in terms of the theoretical hatch-date distribution determination.
	
	The assumption behind the utilization of a constant mortality rate ignores an age or length-specific mortality, which is more likely to realistically occur than a fixed mortality coefficient \citep{Campana:1992wb}. Despite fixed coefficients have been commonly used, as in this work, variable mortality rates along age or size would be a more accurate mathematical approach in order to better obtain the theoretical hatch-date distribution. Variable mortality values could be incorporated in our model as we are simulating the age and size along the entire life period of each individual when the birth date is back-calculated. However, these data are difficult to determinate and are not available yet for this species. 
	This is the first time that a methodology is provided to estimate the theoretical hatch-date distribution from juveniles after the critical mortality stage and already incorporated and sampled from the fishery, accounting for a constant mortality rate. Moreover, our model incorporates the environmental factors with strong influence on the shape of the size distribution due to regional thermal regimes. The thermal regimes are incorporated together with the GSI information of different dolphinfish populations analysed around the globe, as well as the individual daily thermal histories, which are implicit in the growth model used to back-calculate the hatching-date distribution (Moltó et al. submitted).
	
	This work provides a methodology to recover the theoretical hatch-date distribution from a certain cohort. Thus, it can be used to set-up the expected hatch dates distribution under a specific thermal regime; and, due to its coupling to a growth model, used to develop population-based simulations of expected length distribution at a given date, under a given temperature  and management (fishing mortality rates) scenarios.

	\section*{Acknowledgements}
	This work has received funding from the European Union’s Horizon 2020 Research and Innovation program under Grant Agreement 678193 (CERES). AO was supported by H2020 Marie Skłodowska-Curie Actions [746361]. The authors are grateful to Andreina Fenech, Marie Louise Pace and Roberta Mifsud from the Department of Fisheries and Aquaculture of Malta for the data collection.
	
	\bibliography{ArXiv_preprint_aospina}
	
	\appendix
	\label{sec:Appendix_1}
	\clearpage
\onecolumn
\section{Dolphinfish juveniles’ otolith readings database}
\begin{longtable}[c]{lllll}
	\caption{Otolith readings of juvenile dolphinfish, including date of capture, for 13 Mediterranean cohorts (3 years from the Balearic Islands and 8 years from Malta).}
	\label{tab:table-1}\\
	\textbf{Area} & \textbf{Catch Date} & \textbf{Age} & \textbf{FL (cm)} & \textbf{Birthdate}\\
	\endfirsthead
	\multicolumn{5}{c}%
	{{\bfseries Table \thetable\ continued from previous page}} \\
	\textbf{Area} & \textbf{Catch Date} & \textbf{Age} & \textbf{FL (cm)} & \textbf{Birthdate}\\
	\endhead
	Balearic Islands & 09/11/2004 & 97 & 37.5 & 04/08/2004 \\
	Balearic Islands & 09/11/2004 & 97 & 40.4 & 04/08/2004 \\
	Balearic Islands & 05/11/2004 & 95 & 42.6 & 02/08/2004 \\
	Balearic Islands & 09/11/2004 & 170 & 44.9 & 23/05/2004 \\
	Balearic Islands & 05/11/2004 & 139 & 44.6 & 19/06/2004 \\
	Balearic Islands & 09/11/2004 & 112 & 43.1 & 20/07/2004 \\
	Balearic Islands & 05/11/2004 & 135 & 45.0 & 23/06/2004 \\
	Balearic Islands & 05/11/2004 & 131 & 46.6 & 27/06/2004 \\
\end{longtable}
		\label{sec:Appendix_2}
	\clearpage
\onecolumn
\section{IGS databases}
\begin{longtable}[c]{llllllll}
	\caption{Monthly population-averaged GSI for 7 populations around the world (code and source): Taiwan (TAIW, \citealt{Wu:2001es}), Malta (MAL, \citealt{Gatt:2015bh}) , Balearic Islands (BAL, \citealt{Massuti:1995wx} and this work), North Carolina (NCAROL,  \citealt{Schwenke:2008tva}), South West India (SWINDIA, \citealt{Rajesh:2016hh}), Japan (JAP,  \citealt{Furukawa:2012kq}) and California (CALIF, \citealt{ZunigaFlores:2011gz}). The mean, standard deviation and delta Sea Surface Temperature (SST) are indicated.}
	\label{tab:table-2}\\
	\textbf{Area} & \textbf{Month} & \textbf{Mean SST }& \textbf{SD SST} & \textbf{Delta SST} & \textbf{Mean Latitude} & \textbf{Mean IGS} & \textbf{N (IGS)} \\
	\endfirsthead
	\multicolumn{8}{c}%
	{{\bfseries Table \thetable\ continued from previous page}} \\
	\textbf{Area} & \textbf{Month} & \textbf{Mean SST }& \textbf{SD SST} & \textbf{Delta SST} & \textbf{Mean Latitude} & \textbf{Mean IGS} & \textbf{N (IGS)} \\
	\endhead
	TAIW & 1 & 22.752 & 0.502 & 0.063 & 23.350 & 3.713 & NA \\
	TAIW & 2 & 23.242 & 0.614 & 0.490 & 23.350 & 5.090 & NA \\
	TAIW & 3 & 24.515 & 0.546 & 1.273 & 23.350 & 5.150 & NA \\
	TAIW & 4 & 26.445 & 0.404 & 1.930 & 23.350 & 4.132 & NA \\
	TAIW & 5 & 27.714 & 0.480 & 1.269 & 23.350 & 3.952 & NA \\
	TAIW & 6 & 28.717 & 0.340 & 1.003 & 23.350 & 4.012 & NA \\
	TAIW & 7 & 28.759 & 0.387 & 0.042 & 23.350 & 2.814 & NA \\
	TAIW & 8 & 28.281 & 0.309 & -0.478 & 23.350 & 3.533 & NA \\
\end{longtable}
	
\end{document}